\documentstyle[prl,aps,epsf,eqsecnum]{revtex}
\begin{document}
\title{Effective Hamiltonian for 
striped and paired states at the half-filled Landau level} 
\author{Nobuki Maeda}
\address{
Department of Physics, Hokkaido University, 
Sapporo 060-0810, Japan}
\maketitle
\begin{abstract}
We study a pairing mechanism for the quantum Hall system using a mean 
field theory with a basis on the von Neumann lattice, on which the 
magnetic translations commute. 
In the Hartree-Fock-Bogoliubov approximation, we solve the gap equation 
for spin-polarized electrons at the half-filled Landau levels. 
We obtain an effective Hamiltonian which shows a continuous transition 
from the compressible striped state to the paired state. 
Furthermore, a crossover occurs in the pairing phase. 
The energy spectrum and energy gap of the quasiparticle in the paired 
state is calculated numerically at the half-filled second Landau level. 
\end{abstract}
\draft

\ 


\section{Introduction}

The fractional quantum Hall effect (FQHE)\cite{a} is observed around rational 
filling factors with odd denominators in the lowest Landau level. 
This effect is caused by the formation of the incompressible liquid state
\cite{b}. 
Recent experiments showed that the compressible state around the half-filled 
lowest Landau level is similar to the Fermi liquid. 
In the composite fermion theory\cite{c}, this Fermi-liquid-like state is 
described by the composite of an electron and two flux quanta, 
which behaves like a free electron with no magnetic field 
in the mean field approximation. 

In the second Landau level, the situation dramatically changes; 
the FQHE is observed around the half-filling\cite{d}. 
The origin of this quantum Hall state is still mysterious. 
Recent numerical works\cite{e,f} seem to support the Pfaffian state
\cite{g,h,fra}, which has the spin-polarized p-wave pairing potential, 
rather than the spin-singlet paired state\cite{sing}. 
However the microscopic explanation for this pairing mechanism is not 
known yet. 
More quantitative search is, therefore, needed on the basis of the 
microscopic BCS-like model. 
In the present paper, we study the paired state in the mean field theory 
starting from a microscopic Hamiltonian. 

It was shown that the Fermi-liquid-like state\cite{i} can be constructed at 
an arbitrary filling factor in the von Neumann lattice formalism\cite{j}. 
In a strong magnetic field, the free kinetic energy is quenched and the 
kinetic energy is generated from the Coulomb interaction. 
If the translational symmetry on the von Neumann lattice and U(1) symmetry 
are unbroken, the Fermi surface is formed in the magnetic Brillouin zone 
in the mean field theory. 
In the Hartree-Fock approximation, this state corresponds to the striped 
state\cite{k} which is observed at half-filled second and higher Landau 
levels\cite{stripe,o}. 
In Ref.~\cite{f}, the first order transition from striped state to 
paired state is obtained in the numerical calculation of small systems. 
We regard this striped state as the normal state and examine the U(1) 
symmetry breaking mechanism. 
The hopping potential $\varepsilon(p)$ and gap potential $\Delta(p)$ are 
determined 
self-consistently in the Hartree-Fock-Bogoliubov approximation. 

Naively it seems that the formation of the paired state by only the 
repulsive force is impossible in the mean field theory. 
However, screening effect could make the potential attractive 
for the pairing. 
In fact, the quantum effect of fermion loop screens the 
Coulomb potential and the pseudopotential in the Landau level space 
becomes attractive. 
Depending on strength of the screening effect, there are two phases, 
that is, stripe phase and pairing phase. 
On the basis of this observation, we construct an effective Hamiltonian 
which shows a transition from the striped state to the paired state. 
The transition is continuous and very smooth. 
The energy spectrum of the quasiparticle is obtained and energy gap is 
calculated numerically. 
Furthermore, we find a crossover from the pairing phase to the gap-dominant 
pairing phase in which the low-energy excitation occurs around zeros of 
the gap potential. 
 
The paper is organized as follows. 
In Sec. II, a mean field theory on the von Neumann lattice is presented and 
self-consistency equations are obtained. 
In Sec. III, we analyze the self-consistency equations and a screening effect. 
It is shown that there are a stripe phase and pairing phase depending on 
strength of the screening effect. 
In Sec. IV, we present an effective Hamiltonian which 
shows a transition from the striped state to the paired state and calculate 
the energy spectrum of the quasiparticle. 
Summary and discussion are given in Sec. V. 

\section{Hartree-Fock-Bogoliubov approximation}

We consider two-dimensional electron systems in the presence of a 
perpendicular uniform magnetic field $B$. 
The free particle energy is quenched to the Landau level energy 
$E_l=\hbar\omega_c (l+1/2)$, $l=0,1,2\dots$, where $\omega_c=
eB/m_e$ (the cyclotron frequency). 
We suppose that the electrons are spin-polarized and ignore 
the spin degree of freedom. 
In this system, the translation is generated by the magnetic translation 
operators $T(x,y)$ with displacement vector $(x,y)$. 
$T(x,y)$ commutes with the Hamiltonian. 
However, magnetic translation operators are non-commutative, that is, 
\begin{equation}
T(x_1,y_1)T(x_2,y_2)=e^{i{eB\over\hbar}(x_1y_2-x_2y_1)}T(x_2,y_2)T(x_1,y_1).
\end{equation} 
Therefore, the translation becomes commutative if we restrict the 
displacement vectors on a two-dimensional lattice which has a unit cell 
with an area $2\pi\hbar/eB$. 
This lattice is called the von Neumann lattice. 
We consider lattice sites $a(m,n)$, where $m$, $n$ are integers and 
lattice spacing $a=\sqrt{2\pi\hbar/eB}$. 
Then $T(ma,na)$'s commute with each other and have simultaneous 
eigenstates as 
\begin{equation}
T(ma,na)u_{l,p}({\bf r})=e^{i(p_x m+p_y n)a/\hbar}u_{l,p}({\bf r})
\end{equation}
in the $l$ th Landau level. 
We call ${\bf p}=(p_x,p_y)$ the momentum and $u_{l,p}({\bf r})$ the 
Bloch wave on the von Neumann lattice, 
which is defined on the Brillouin zone $\vert p_x\vert$, 
$\vert p_y\vert<\pi\hbar/a$. 
It was proved that the eigenstates form an orthogonal complete set
\cite{pere,j}. 
For simplicity, we set $a=\hbar=c=1$. 

The Bloch wave $u_{l,p}({\bf r})$ is constructed by using 
the coherent states of the guiding center coordinates $(X,Y)$ defined by 
\begin{eqnarray}
X&=&x-\xi,\nonumber\\
Y&=&y-\eta,\\
\xi&=&{1\over eB}(-i{\partial\over\partial y}+eA_y),\nonumber\\
\eta&=&-{1\over eB}(-i{\partial\over\partial x}+eA_x),\nonumber
\end{eqnarray}
where $\nabla\times A=B$ and $(\xi,\eta)$ are called the relative 
coordinates. 
These operators satisfy the following commutation relations, 
\begin{eqnarray}
&[\xi,\eta]=-[X,Y]=-{i\over eB},\\
&[\xi,X]=[\xi,Y]=[\eta,X]=[\eta,Y]=0.\nonumber
\end{eqnarray}
The coherent states of $(X,Y)$ are defined by
\begin{eqnarray}
(X+iY)f_{lmn}({\bf r})=(m+in)f_{lmn}({\bf r}),\\
{m_e\omega^2_c\over2}(\xi^2+\eta^2)f_{lmn}({\bf r})=E_l f_{lmn}({\bf r}). 
\end{eqnarray}
Note that these coherent states are non-orthogonal complete set as
\begin{equation}
\int d^2 r f_{l,m'+m,n'+n}({\bf r})f_{l',m',n'}({\bf r})=
\delta_{ll'}e^{i\pi(m+n+mn)-{\pi\over2}(m^2+n^2)}.
\end{equation}
Using $f_{lmn}({\bf r})$, the orthogonal basis $u_{l,p}({\bf r})$ is 
given by
\begin{equation}
u_{l,p}({\bf r})={1\over\beta(p)}\sum_{mn}e^{ip_x m+ip_y n} 
f_{lmn}({\bf r}),
\end{equation}
where $\beta(p)$ is a normalization factor\cite{j}. 
In this basis, two-dimensional momentum is a good quantum number and 
the Fermi surface is formed in the mean field theory\cite{i}. 
It was shown that the mean field state could explain the anisotropic 
compressible state observed at the half-filled second and higher Landau 
levels\cite{k}. 
Therefore the basis on the von Neumann lattice is useful tool to 
investigate the physics at the half-filled second and higher Landau 
levels. 

Fourier transforming the Bloch wave, we can obtain the orthogonal 
localized basis\cite{loc} 
\begin{equation}
w_{l,{\bf X}}({\bf r})=\int_{\rm BZ}{d^2p\over(2\pi)^2} 
u_{l,p}({\bf r})e^{-i(p_x m+p_y n)+i\lambda(p)}
\label{lam}
\end{equation}
where ${\bf X}=(m,n)$, BZ stands for the Brillouin zone, 
and $\lambda(p)$ represents the gauge degree of freedom which is taken to 
satisfy the periodic condition in the Brillouin zone. 
We call this basis the Wannier basis. 
This basis is localized on a position of the lattice site. 
The explicit forms of these bases are given in Ref.~\cite{j}. 

Using the Bloch wave basis and Wannier basis, we can expand the electron 
field operator $\psi({\bf r})$ as
\begin{eqnarray}
\psi({\bf r})&=&\sum_{l=0}^\infty\int_{\rm BZ}{d^2p\over(2\pi)^2}a_l(p)
u_{l,p}({\bf r})\nonumber\\
&=&\sum_{l=0}^\infty \sum_{\bf X}b_{l}({\bf X}) 
w_{l,{\bf X}}({\bf r}),
\end{eqnarray}
where $a_l(p)$ and $b_l({\bf X})$ are anti-commuting annihilation 
operators in the momentum space and lattice space, respectively. 
The total Hamiltonian of the present system is 
\begin{equation}
H=\int d^2r\psi^\dagger({\bf r}){(-i\nabla+eA)^2\over2m_e}\psi({\bf r})+
{1\over2}\int d^2rd^2r':(\rho({\bf r})-\rho_0)V({\bf r}-{\bf r}')
(\rho({\bf r}')-\rho_0):,
\end{equation}
where colons mean the normal ordering, 
$\rho({\bf r})=\psi^\dagger({\bf r})\psi({\bf r})$, 
$V({\bf r})=q^2/r$, and $\rho_0$ is the uniform background density. 

Let us project the system to the $l$ th Landau level space. 
Up to the Landau level energy $E_l$, the projected Hamiltonian $H^{(l)}$ is 
written as
\begin{equation}
H^{(l)}={1\over2}\sum_{X_1 {X_1}'X_2 {X_2}'}
:b^\dagger_l({\bf X}_1)b_l({\bf X}'_1)V_l({\bf X},{\bf Y},{\bf Z})
b^\dagger_l({\bf X}_2)b_l({\bf X}'_2):,
\label{pro}
\end{equation}
where ${\bf X}={\bf X}_1-{\bf X}'_1$, 
${\bf Y}={\bf X}_2-{\bf X}'_2$, ${\bf Z}={\bf X}_1-{\bf X}'_2$, 
and $V_l({\bf X},{\bf Y},{\bf Z})$ is given as
\begin{eqnarray}
V_l({\bf X},{\bf Y},{\bf Z})&=&\int{d^2k\over(2 \pi)^2}\int_{\rm BZ}
{d^2p_1\over(2 \pi)^2}{d^2p_2\over(2 \pi)^2}{\tilde v}_l(k)
e^{i[p_1\cdot X+p_2\cdot Y+k\cdot Z+f(p_1,p_1+k)+f(p_2+k,p_2)]},\\
f(p+k,p)&=&\int_{p}^{p+k}(\alpha(p')+\nabla_{p'}\lambda(p'))dp'.
\end{eqnarray}
Here $\alpha(p)$ is a vector potential, $\nabla_p=(
{\partial\over\partial p_x},{\partial\over\partial p_y})$, 
$\nabla_p\times\alpha(p)=-1/2\pi$ which is a unit flux penetrating the 
momentum space, and the line integral is along a straight line. 
The effective Coulomb potential in the $l$ th Landau level 
${\tilde v}_l$ is given by
\begin{equation}
{\tilde v}_l(k)=\{L_l({k^2\over4\pi})\}^2e^{-{k^2\over4\pi}}
{2 \pi q^2\over k}, 
\end{equation}
and ${\tilde v}_l(0)=0$ due to the charge neutrality condition, 
where $L_l$ is the Laguerre polynomial. 

The system is translationally invariant in the lattice space and 
in the momentum space with the uniform magnetic field 
$\nabla\times\alpha(p)$. 
The translational symmetry in the momentum space is called the K-invariance. 
We consider the case that 
the K-invariance is spontaneously broken and the kinetic term is 
induced through the correlation effect. 

We apply the mean field approximation to the projected Hamiltonian 
(\ref{pro}). 
Let us consider the following mean fields, 
which are translationally invariant on the von Neumann lattice, 
\begin{eqnarray}
U_l({\bf X}-{\bf X}')&=&\langle b_l^\dagger({\bf X}')b_l({\bf X})\rangle,
\nonumber\\
U_l^{(+)}({\bf X}-{\bf X}')&=&\langle b_l^\dagger({\bf X}')b_l^\dagger
({\bf X})\rangle,\\
U_l^{(-)}({\bf X}-{\bf X}')&=&\langle b_l({\bf X}')b_l({\bf X})\rangle.
\nonumber
\end{eqnarray}
These mean fields break the K-invariance and U(1) symmetry. 
$U^{(\pm)}_l$ satisfies $U_l^{(-)*}({\bf X})=U_l^{(+)}(-{\bf X})$ and 
$U_l^{(-)}(-{\bf X})=-U_l^{(-)}({\bf X})$. 
Using these mean fields, we approximate $H^{(l)}$ as 
\begin{eqnarray}
H^{(l)}_{\rm mean}&=&\sum_{XX'} 
U_l({\bf X}-{\bf X}')\{v_l^H({\bf X}'-{\bf X}')-
v^F_l({\bf X}'-{\bf X})\}b^\dagger({\bf X})b({\bf X}')
\nonumber\\
&&+{1\over2}\sum_{XX'Y} \{U_l^{(+)}({\bf Y})V_l^B({\bf X}-{\bf X}',{\bf Y})
b_l({\bf X})b_l({\bf X}')
\nonumber\\
&&\phantom{+{1\over2}\sum_{XX'Y} \{}
+U_l^{(-)}({\bf Y})V_l^B(-{\bf Y},{\bf X}'-{\bf X})
b_l^\dagger({\bf X})b_l^\dagger({\bf X}')
\}.
\end{eqnarray}
Here the Hartree potential $v^H_l$ and Fock potential $v^F_l$ 
are given by the following sum rules 
\begin{eqnarray}
\sum_Z V_l({\bf X},{\bf Y},{\bf Z})&=&v^H_l({\bf X})\delta^{\bf X}_{-\bf Y},\\
\sum_X V_l({\bf X},{\bf Y}-{\bf X},{\bf Z})&=&
v^F_l({\bf Z})\delta^{\bf Y}_{0},
\end{eqnarray}
where
\begin{eqnarray}
v^H_l({\bf X})&=&{\tilde v}_l(2 \pi{\bf X}),\\
v^F_l({\bf X})&=&\int{d^2 k\over (2\pi)^2}{\tilde v}_l(k)e^{ik\cdot X}.
\label{fo}
\end{eqnarray}
We define the Hartree-Fock potential as $v_l^{\rm HF}({\bf X})=
v_l^H({\bf X})-v_l^F({\bf X})$ and 
its Fourier transform as ${\tilde v}_l^{\rm HF}(p)$. 
The Bogoliubov potential $V_l^B$ is written as 
\begin{eqnarray}
V^B_l({\bf X},{\bf Y})&=&\sum_{Z}V_l({\bf X}+{\bf Z},{\bf Y}+{\bf Z},
{\bf Z})\nonumber\\
&=&\int_{\rm BZ}{d^2 p_1\over (2\pi)^2}{d^2 p_2\over (2\pi)^2}
{\tilde V}^B_l(p_1,p_2)e^{ip_1\cdot X+ip_2\cdot Y},
\end{eqnarray}
and ${\tilde V}^B_l(p_1,p_2)$ is given by
\begin{equation}
{\tilde V}^B_l(p_1,p_2)=\sum_N {\tilde v}_l(p_1+p_2+2 \pi N) 
e^{i[f(p_1,-p_2-2 \pi N)+f(-p_1,p_2+2 \pi N)]},
\label{bo}
\end{equation}
where $N=(N_x,N_y)$, $N_x$, $N_y$ are integers. 
Then we introduce the hopping potential $\varepsilon$ 
and gap potential $\Delta$ as
\begin{eqnarray}
\varepsilon_{l,{\bf X}}&=&U_l({\bf X}) v_l^{HF}(-{\bf X}),
\label{ep}\\
\Delta_{l,{\bf X}}&=&\sum_Y 
U^{(-)}_l({\bf Y}) V_l^{B}(-{\bf X},-{\bf Y}).
\label{de}
\end{eqnarray}
We define these Fourier transforms as $\varepsilon_l(p)$ and $\Delta_l(p)
e^{-i(\lambda(p)+\lambda(-p))}$, respectively. 
The phase factor of the gap potential is the same in Eq.~(\ref{lam}). 
$V^B_l$ satisfies $V^B_l(-{\bf X},-{\bf Y})=V^B_l({\bf X},{\bf Y})$ and 
we have $\Delta_{l,-X}=-\Delta_{l,X}$. 
Using these potentials, the mean field Hamiltonian reads
\begin{equation}
H^{(l)}_{\rm mean}=\sum_{XX'}[\varepsilon_{l,{\bf X}-{\bf X}'}
b_l^\dagger({\bf X})b_l({\bf X}')+
{1\over 2}\Delta^*_{l,{\bf X}'-{\bf X}}b({\bf X})b({\bf X}')+
{1\over 2}\Delta_{l,{\bf X}'-{\bf X}}b^\dagger({\bf X}')b^\dagger({\bf X})].
\label{mham}
\end{equation}
This Hamiltonian is diagonalized by the Bogoliubov transformation in the 
momentum space. 
We fix the gauge of vector potential $\alpha$ in the momentum space as
$\alpha(p)=(p_y/2\pi,0)$. 
In this gauge, the boundary conditions in the Brillouin zone are given by
\begin{eqnarray}
a_l(p+2 \pi N)&=&e^{i\pi(N_x+N_y)-iN_yp_x}a_l(p),\\
\Delta_l(p+2\pi N)&=&e^{-2iN_yp_x}\Delta_l(p).
\label{boun}
\end{eqnarray}
Equation (\ref{boun}) and the relation $\Delta_l(-p)=-\Delta_l(p)$ make 
the gap potential $\Delta_l(p)$ have four zeros at ${\bf p}=(0,0)$, 
$(0,\pi)$, $(\pi,0)$, and $(\pi,\pi)$. 
We assume $\varepsilon_l(-p)=\varepsilon_l(p)$. 
$U_l$ and $U_l^{(-)}$ are calculated by using $H^{(l)}_{\rm mean}$. 
Then, the hopping potential and gap potential in Eqs.~(\ref{ep}) and 
(\ref{de}) are determined by the self-consistency equations, 
\begin{eqnarray}
\xi_l(p)&=&\int_{\bf BZ}{d^2p_1\over(2\pi)^2}{E_l(p_1)-\xi_l(p_1)\over2
E_l(p_1)}{\tilde v}_l^{\rm HF}(p_1-p)-\mu,\label{selfx}\\
\Delta_l(p)&=&-\int{d^2p_1\over(2\pi)^2}{\Delta_l(p_1)\over2
E_l(p_1)}{\tilde v}_l(p_1-p)e^{i\int_p^{p_1}
2\alpha(k)dk},
\label{self}
\end{eqnarray}
where $\mu$ is the chemical potential and $E_l(p)$ is the 
spectrum of the quasiparticle which is defined by
\begin{eqnarray}
E_l(p)&=&\sqrt{\xi_l(p)^2+\vert\Delta_l(p)\vert^2},\\
\xi_l(p)&=&\varepsilon_l(p)-\mu.
\end{eqnarray}
These self-consistency equations are equivalent to 
sum up infinite diagrams of the Fermion self-energy part. 
Note that the integral region of Eq.~(\ref{self}) are infinite. 
Note also that the gauge field appears in the gap equation (\ref{self}). 
The gauge field represents two unit flux on the Brillouin zone. 
This corresponds to the flux carried by the Cooper pair $\langle a(p)
a(-p)\rangle$ which is separated between $p$ and $-p$ in the Brillouin zone. 
The mean field $\langle a^\dagger(p)a(p)\rangle$, on the other hand, is 
placed on the same point. 
Therefore no gauge field appears in Eq.~(\ref{selfx}). 

The filling factor of the $l$ th Landau level is denoted by $\nu_l$ and 
total filling factor is given by $\nu=l+\nu_l$. 
Using the mean fields, the filling factor is given by 
\begin{equation}
\nu_l={1\over2}-\int_{\bf BZ}{d^2p\over(2\pi)^2}{\xi_l(p)\over 2E_l(p)}. 
\label{nu}
\end{equation}
In the limit of $\Delta\rightarrow0$, $\nu_l$ is equal to $\int_{\bf BZ}
\theta(\mu-\varepsilon_l(p))d^2p/(2\pi)^2$.

\section{Stripe phase and pairing phase}

In this section, we analyze the self-consistency equations (\ref{selfx}) 
and (\ref{self}). 
It is shown that the screening effect plays important roles 
for the pairing mechanism. 
For simplicity, we omit the Landau level index $l$ and use $q^2/a$ as 
the unit of energy. 

It is convenient for solving Eq.~(\ref{selfx}) to use the following 
mode expansions, 
\begin{eqnarray}
\xi(p)&=&\sum_X \xi_{\bf X} e^{iX\cdot p},
\label{exxi}\\
{\xi(p)\over E(p)}&=&\sum_X \eta_{\bf X} e^{iX\cdot p},
\end{eqnarray}
where $\xi_{-\bf X}=\xi_{\bf X}$, $\eta_{-\bf X}=\eta_{\bf X}$, and 
$\xi_{\bf X}$, $\eta_{\bf X}$ are real numbers. 
Then Eq.~(\ref{selfx}) becomes
\begin{eqnarray}
\xi_{\bf X}&=&-{v^{\rm HF}({\bf X})\over2}\eta_{\bf X},{\ \rm for\ }{\bf X}
\neq0,\\
\xi_{\bf 0}&=&-{v^{\rm HF}({\bf 0})\over2}(\eta_{\bf 0}-1)-\mu.
\label{exxixi}
\end{eqnarray}
The zero mode $\eta_{\bf 0}$ is related to the filling factor as 
$\nu_l=(1-\eta_{\bf 0})/2$ by Eq.~(\ref{nu}). 
Therefore the zero mode $\xi_{\bf 0}$ is determined by $\mu$ and $\nu_l$. 
In particular, $\eta_{\bf 0}=0$ for the half-filling case ($\nu_l=1/2$). 
We introduce hopping parameters $t_{\bf X}=-2\xi_{\bf X}$ and regard 
$\xi_{\bf X}$ as a functional of $t_{\bf X}$'s and $\Delta(p)$. 
Then the self-consistency for $t_{\bf X}$ is written as 
\begin{equation}
t_{\bf X}=-2 \xi_{\bf X}[\{t_{(m,n)}\},\Delta(p)]. 
\end{equation}
Using $t_{(m,n)}$, the hopping term is written by $-\sum_{m,n\geq0} 
t_{(m,n)} \cos(p_x m+p_y n)a^\dagger(p) a(p)$. 

Next we discuss the self-consistency for the gap potential $\Delta(p)$. 
Eq.~(\ref{self}) is rewritten as 
\begin{equation}
\Delta(p)=-\int{d^2k\over(2\pi)^2}{\tilde v}(k) e^{ik\cdot D}{\Delta(p)
\over 2E(p)},
\end{equation}
where ${\bf D}=(-i{\partial\over\partial p_x}+2\alpha_x,
-i{\partial\over\partial p_y}+2\alpha_y)$. 
It is convenient to introduce eigenfunctions of the operator ${\bf D}^2$, 
that is, ${\bf D}^2 \psi_n(p)=e_n \psi_n(p)$ with $e_n=(2 n+1)/\pi$, 
$n=0,1,2,\dots$. 
The index $n$ labels the $n$ th Landau level in the momentum space. 
The eigenfunctions which obey the same boundary condition as 
the gap function are doubly degenerate and are given by
\begin{eqnarray}
\psi^{(2)}_n(p)&=&N_n \sum_{N_y} H_n({p_y+(2 N_y+1)\pi\over\sqrt{2\pi}})
e^{i(2 N_y+1)p_x-{1\over2\pi}(p_y+(2 N_y+1)\pi)^2}
\nonumber\\
&=&{2\pi\over\sqrt{n!}}({\pi\over2})^{n/2}(D_x-iD_y)^n
\psi^{(2)}_0(p),\\
\psi^{(3)}_n(p)&=&N_n \sum_{N_y} H_n({p_y+2 N_y\pi\over\sqrt{2\pi}})
e^{i 2 N_y p_x-{1\over2\pi}(p_y+2 N_y \pi)^2}
\nonumber\\
&=&{2\pi\over\sqrt{n!}}({\pi\over2})^{n/2}(D_x-iD_y)^n \psi^{(3)}_0(p),
\end{eqnarray}
where $H_n$ is the Hermite polynomial, $N_n=1/\sqrt{2^{n-1}n!}$, and 
$\psi^{(2)}_0$, $\psi^{(3)}_0$ are written by theta functions as 
\begin{eqnarray}
\psi^{(2)}_0(p)&=&N_0 e^{-{p_y^2\over2\pi}}
\vartheta_2({p_x+ip_y\over\pi}\vert 2i),
\label{ptwo}\\
\psi^{(3)}_0(p)&=&N_0 e^{-{p_y^2\over2\pi}}
\vartheta_3({p_x+ip_y\over\pi}\vert 2i).
\label{pthree}
\end{eqnarray}
These eigenfunctions have the parity symmetry, $\psi_n(-p)=(-)^n\psi_n(p)$. 
Therefore the gap potential can be expanded by $\psi_{2 n+1}(p)$. 
Furthermore the eigenfunctions satisfy the following relations 
\begin{eqnarray}
\psi^{(2)}_n(p_x+\pi,p_y)&=&-\psi^{(2)}_n(p_x,p_y),\\
\psi^{(3)}_n(p_x+\pi,p_y)&=&\psi^{(3)}_n(p_x,p_y),\\
\psi^{(2)}_n(p_x,p_y+\pi)&=&e^{-ip_x}\psi^{(3)}_n(p_x,p_y),\\
\psi^{(3)}_n(p_x,p_y+\pi)&=&e^{-ip_x}\psi^{(2)}_n(p_x,p_y),
\end{eqnarray}
and $\psi_n(\pi m,\pi n)=0$. 
Moreover $\psi^{(2)}_n(\pi n/2,\pi)=0$ and $\psi^{(3)}_n(\pi n/2,0)=0$. 
We can expand the gap potential as
\begin{eqnarray}
\Delta(p)&=&\sum_{n\geq1,i=2,3}c_n^{(i)}\psi_{2 n-1}^{(i)}(p),
\label{exde}\\
{\Delta(p)\over E(p)}&=&\sum_{n\geq1,i=2,3}d_n^{(i)}\psi_{2 n-1}^{(i)}(p).
\end{eqnarray}
Then Eq.~(\ref{self}) becomes
\begin{eqnarray}
c_n^{(i)}&=&-{F_{2 n+1}\over2}d_n^{(i)},\\
F_n&=&\int{d^2 k\over(2\pi)^2}{\tilde v}(k)L_n({k^2\over2\pi})
e^{-{k^2\over4\pi}}.
\label{exdede}
\end{eqnarray}
Note that although $F_n$'s have the same form as the 
Haldane's pseudopotentials\cite{hal}, the physical meaning is different. 
Haldane's one is the 2-body interaction projected to the relative angular 
momentum in the single Landau level. 
It is not screened by definition. 
The potential $F_n$, on the other hand, is a potential appearing in 
the gap equation which is renormalized by higher order corrections. 
Actually the potential $F_n$ is screened by the polarization 
$\Pi(p)$ due to the Fermion loop diagrams. 
We approximate the screened potential as
\begin{eqnarray}
{\tilde v}(p,m_{\rm TF})&=&1/({\tilde v}(p)^{-1}+m_{\rm TF}),\\
m_{\rm TF}&=&-\Pi(0),
\end{eqnarray}
where $m_{\rm TF}$ is the Thomas-Fermi mass. 
Calculating the one-loop diagram, $m_{\rm TF}$ is given by 
\begin{equation}
m_{\rm TF}=\int_{\bf BZ}{d^2p\over(2\pi)^2}{\vert\Delta(p)\vert^2
\over 2E(p)^3}
\label{tf}
\end{equation}
We use the screened potential ${\tilde v}(p,m_{\rm TF})$ in the Fock 
potential (\ref{fo}) and Bogoliubov potential (\ref{bo}). 
In the zero limit of the gap potential $\Delta$, $m_{\rm TF}$ 
becomes the density of states on the Fermi surface. 
It should be noted that the Hartree potential $v^H$ is not 
screened because the polarization effect is automatically 
included in the self-consistency equation. 

The $m_{\rm TF}$ dependences of the hopping strength $v^{\rm HF}({\bf X})$ 
and pseudopotentials $F_n$ are plotted in Figs.~(1)-(6) for $l=$0, 1, 2. 
As seen in these figures, potentials change their signs at 
$m_{\rm TF}\approx 1.0\sim2.0$. 
Especially for $l=$1 and 2, all potentials change their signs together 
at large $m_{\rm TF}$. 
This observation leads us to a strong statement for $l=1$ and 2. 
Using Eqs.~(\ref{exxi})-(\ref{exxixi}) and Eqs.~(\ref{exde})-
(\ref{exdede}), the following inequalities are concluded at $\nu_l=1/2$, 
\begin{eqnarray}
\int{\xi(p)^2\over E(p)}{d^2k\over(2\pi)^2}&=&
-\sum_{X\neq0}{v^{\rm HF}({\bf X})\over2}\eta^2_{\bf X}\geq0,\\
\int{\vert\Delta(p)\vert^2\over E(p)}{d^2k\over(2\pi)^2}&=&
-\sum_{n\geq1,i=2,3}{F_n\over2}\vert d^{(i)}_n\vert^2\geq0.
\end{eqnarray}
Therefore, $d_n^{(i)}=0$ for all $n$ and $i$ at small $m_{\rm TF}$ 
and we obtain U(1) symmetric phase ($\Delta(p)=0$). 
This phase was studied in the Hartree-Fock approximation. 
It was shown that the compressible striped state is favored in this 
phase\cite{m,n,k}. 
At large $m_{\rm TF}$ for $l=$1 and 2, $\eta_{\bf X}=0$ for 
${\bf X}\neq0$ and we obtain a pairing phase. 
Thus, in the Hartree-Fock-Bogoliubov approximation, 
the stripe phase makes transition to the pairing phase 
for $m_{\rm TF}>1.0\sim 2.0$. 
In this pairing phase, $\xi=0$ and the energy spectrum of the quasiparticle 
becomes $\vert\Delta(p)\vert$. 
Hence, there are at least four gapless points in 
the Brillouin zone for the obtained paired state. 
This peculiar result conflicts with the experiment\cite{d} and numerical 
calculations\cite{e,f}, in which the paired state has an energy gap. 
Furthermore we find that the self-consistency for $m_{\rm TF}$ 
(\ref{tf}) is not satisfied, which goes to infinity as iterating 
the numerical calculations. 
This means that the fluctuation is too large and the mean 
field solution is not stable. 
In the next section, we change the summation of infinite  
diagrams of the Fermion self-energy part and obtain 
a self-consistent paired state with an energy gap. 

\section{Effective Hamiltonian for the striped and paired state}

In this section, we present an effective Hamiltonian which 
has a striped state as the normal state and paired state as the 
U(1) symmetry breaking state. 
We focus our argument on the half-filled second Landau level 
space, that is $l=1$, $\nu=1+1/2$. 
We use a tricky technique to find a self-consistent solution for 
the paired state with an energy gap. 
The most hardest problem in the quantum Hall system is to derive  
a low-energy effective theory from the microscopic Hamiltonian, 
because we do not know a small parameter used in the 
perturbation expansion a priori. 
Therefore we have to choose a starting point by consideration of 
symmetry and physical intuition based on experiments. 
In the present case, we maintain the translational symmetry on 
the von Neumann lattice and choose the striped state as the normal 
state. 

Let us divide the mean field Hamiltonian (\ref{mham}) into two parts as
\begin{eqnarray}
H_{\rm mean}&=&H_0+H_1,\\
H_1&=&\int_{\rm BZ}{d^2p\over(2\pi)^2}t\cos p_y a^\dagger(p)a(p), 
\end{eqnarray}
and $H_1$ is treated as perturbation to $H_0=-H_1+H_{\rm mean}$. 
Self-consistency is imposed for $H_0$ first and correction from 
$H_1$ is included next. 
$H_1$ represents the anisotropy in the stripe state or normal state 
in which the uniform direction is chosen in the $y$ direction. 
In the striped state at the half-filling, the Fermi sea is formed at 
$\vert p_y\vert < \pi/2$, which is shown in Ref.~\cite{k} 
In this case the Fermi surface (line) is formed at $p_y=\pm\pi/2$. 
It is expected that the paired state possesses the same anisotropy as 
the normal state. 
In $H_0$, the nearest-neighbor (N-N) hopping parameter becomes 
$t-2\xi_{(0,1)}=t+t_{(0,1)}$. 
Then the self-consistent hopping parameter satisfies 
\begin{equation}
t_{\bf X}=-2 \xi_{\bf X}[\{t\delta_{m,n}^{0,1}+t_{(m,n)}\},\Delta(p)]. 
\end{equation}
The N-N hopping parameter is renormalized as $t'=t+t_{(0,1)}$ in $H_0$. 
Including the first order correction of $H_1$, the effective hopping 
parameter becomes $t_{\rm eff}=t'-t=t_{(0,1)}$. 
The parameter $t$ controls the effective N-N hopping strength dynamically. 
The magnitude of $t_{\rm eff}$ corresponds to the strength of the stripe 
order. 

We truncate the expansions of Eqs.~(\ref{exxi}) and (\ref{exde}) and 
calculate the self-consistent solution by iteration of numerical 
calculations until we obtain convergence. 
Considering only the lowest and next relevant terms, the effective 
Hamiltonian for the quasiparticle in the striped and paired state 
is given by
\begin{equation}
H_{\rm eff}=\int_{\rm BZ}{d^2p\over(2\pi)^2}[
\varepsilon_{\rm eff}(p)
a^\dagger(p) a(p)+{1\over2}\Delta_{\rm eff}(p) 
a^\dagger(-p) a^\dagger(p)+{1\over2}\Delta^*_{\rm eff}(p) a(p)a(-p)],
\end{equation}
where the effective hopping potential and effective gap potential 
are given by 
\begin{eqnarray}
\varepsilon_{\rm eff}(p)&=&-t_{\rm eff}\cos p_y-t_{(0,3)}\cos 3 p_y,
\label{epe}\\
\Delta_{\rm eff}(p)&=&c_1^{(3)}\psi_1^{(3)}(p)+c_3^{(3)}\psi_3^{(3)}(p). 
\label{dele}
\end{eqnarray}
In using Eqs.~(\ref{epe}) and (\ref{dele}), 
$\psi_n^{(2)}$ and $\cos((2m+1)p_x+np_y)$ terms are irrelevant and 
$\cos(2m p_x+2 n p_y)$ terms induced in $\varepsilon_{\rm eff}$ 
give a negligible correction to the numerical results. 
The $t$ dependence of $t_{\rm eff}$ is determined self-consistently by 
$t_{\rm eff}=-2 \xi_{(0,1)}[t+t_{\rm eff},t_{(0,3)},c_1^{(3)},
c_3^{(3)}]$. 
The $t$ dependence is implicit in the effective theory. 
The hopping parameters $t_{(0,n)}$ and gap potential $\Delta_{\rm eff}(p)$ 
depend on $t_{\rm eff}$ explicitly. 
We use the following approximation 
\begin{equation}
m_{\rm TF}={1\over\pi t_{\rm eff}},
\label{tfa}
\end{equation}
which is satisfied in the stripe phase with only the N-N hopping term. 
We checked that Eq.~(\ref{tfa}) was good approximation using the obtained 
convergent solutions. 

Chemical potential $\mu$ is determined so that the filling factor 
$\nu_1$ is equal to $1/2$. 
Note that $\mu$ includes the on-site term in $H_{\rm mean}$. 
We find that $\mu$ is negative small number on the order of $10^{-4}$ 
at most. 
The gap potential $\Delta_{\rm eff}(p)$ depends on $t_{\rm eff}$. 
Figure (7) shows the $t_{\rm eff}$ dependence of 
the energy gap $\Delta E=\min(2 E(p))$ at $\nu=1+1/2$. 
The maximum value of the energy gap is $0.027$ at $t_{\rm eff}=0.03$. 
In this case, $t_{(0,3)}=-0.0003$, $c_1^{(3)}=0.0104$, $c_3^{(3)}=-0.0018$, 
and $\mu=-0.0006$. 
The energy spectrum of the quasiparticle at $t_{\rm eff}=0.03$ 
is shown in Fig.~(8). 
The absolute value of the gap potential at $t_{\rm eff}=0.03$ is shown 
in Fig.~(9). 
The excitation energy $E(p)$ becomes small around $p_y=\pm\pi/2$, 
which is the Fermi surface of the striped state. 
The transition to the stripe phase is continuous and very smooth. 
Near the transition point, the behavior of energy gap is approximated by  
$t_{\rm eff} e^{-2\pi t_{\rm eff}/\vert F_1\vert}$, whose non-perturbative 
dependence on the coupling is well-known in the BCS theory. 
At $m_{\rm TF}= m_c\approx 1.4$, $F_1$ behaves as $\alpha (m_c-m_{\rm TF})$ 
as seen in Fig.~(4), and the energy gap approaches to zero as
\begin{equation}
\Delta E\propto t_{\rm eff} e^{-{2\pi t_{\rm eff}^2\over \alpha m_c 
(t_c-t_{\rm eff})}},\quad {\rm for}\  t_{\rm eff}<t_c, 
\end{equation}
where $t_c\approx 1/\pi m_c\approx0.2$. 
The energy gap is extremely small at $0.1<t_{\rm eff}<t_c$. 
At $t_{\rm eff}>t_c$, the gap potential vanishes and the compressible 
striped state is realized. 
The spectrum of the striped state is uniform in the $p_x$ direction. 
This state is regarded as a collection of the one-dimensional lattice 
Fermion systems and leads to the anisotropy of the magnetoresistance
\cite{k}. 

Inspecting the energy spectrum of quasiparticle in the pairing phase, 
we find a crossover phenomenon at $t_{\rm eff}\sim0.01$, 
that is, the minimum excitation energy $\min(E(p))$ is placed around 
$p_y=\pm\pi/2$ at $0.01<t_{\rm eff}<t_c$, 
whereas at $0<t_{\rm eff}<0.01$, placed around $p_y=0$ and 
$\pi$. 
We call the latter case the gap-dominant pairing phase. 
In this phase, the low-energy excitation occurs around zeros of the 
gap potential and the the energy gap is given by $2\xi(0)$. 
Therefore the $t_{\rm eff}$ dependence of energy gap $\Delta E$ 
becomes linear at $0<t_{\rm eff}<0.01$
The energy spectrum at $t_{\rm eff}\sim0.01$ is shown in Fig.~(10). 
As seen in this figure, the spectrum is close to the flatband and 
the stripe order is weakened compared with Fig.~(8). 

\section{Summary and discussion}

We propose an effective Hamiltonian which describes the striped and 
paired state at the half-filled Landau levels. 
The gap potential and screening mass are determined self-consistently 
in the Hartree-Fock-Bogoliubov approximation scheme. 
The energy spectrum of the quasiparticle and energy gap are 
calculated numerically. 

The very smooth transition from pairing to stripe phase occurs at 
$t_{\rm eff}=t_c$. 
In other words, the stripe order strongly suppresses the pairing order 
at $0.1<t_{\rm eff}<t_c$. 
The energy gap becomes maximum at $t_{\rm eff}=0.03$ and the 
maximum value is $0.027q^2/a=0.01q^2/l_B$ ($l_B=\sqrt{\hbar/eB}$) 
which is the same order as Morf's value\cite{e}. 
The experimental value at $\nu=5/2$ is smaller than the theoretical 
one by an order of magnitude\cite{d}. 
This difference could be explained by effects of disorder and 
finite thickness of 2D layer, which reduce the energy gap
\cite{will}. 
The regions of small $t_{\rm eff}$ and $0.1<t_{\rm eff}<t_c$ have 
another possibility to explain the experiment. 
In our model, the crossover is obtained for  small $t_{\rm eff}$. 
This suggests that the stripe order is destroyed by the pairing order 
and the system becomes isotropic in the gap-dominant pairing phase. 

The electric charge of the quasiparticle is a half of the 
electron charge and the Hall conductance is quantized as $e^2/2h$. 
The results are consistent with the recent experiment. 
However, we do not know whether correction to the quantized Hall 
conductance is finite or zero. 
Related to this subject, $U(1)$ breaking effect in the $PT$ violating 
system is studied in \cite{q}. 
Whether the Hall conductance in the paired state is topological 
invariant or not is an interesting future problem. 

It is possible to consider a pairing mechanism based on the composite 
fermion picture\cite{l}. 
However, it is unclear that the composite fermion picture is applicable to 
the second and higher Landau levels. 
It rather seems that the compressible stripe state\cite{m} which has an 
anisotropic Fermi surface is the normal state at the half-filled second 
and higher Landau levels in the Hartree-Fock approximation and 
numerical calculations of small systems\cite{n,k}. 
In fact, the transition from the paired state to striped state is observed 
in the presence of an inplane magnetic field\cite{o}. 
Furthermore Ref.~\cite{bon} suggests that the Chern-Simons gauge 
fluctuations are strongly pair breaking. 
Controversially, a superfluid state at the half-filling Hall state is 
proposed\cite{bas} on the basis of the dipolar liquid picture of the 
composite fermion\cite{dip}. 

Relation between our model and the Pfaffian state is not understood yet. 
Our effective Hamiltonian for the striped and paired state resembles the 
effective Hamiltonian proposed by Read and Green\cite{p}, which has the 
Pfaffian state in the weak pairing phase. 
However their normal state has the isotropic Fermi surface, 
which is reminiscent of the composite Fermion theory, and is different 
from our anisotropic one. 
The p-wave behavior seen in Eqs.~(\ref{ptwo}) and (\ref{pthree}), 
on the other hand, is in common with the Pfaffian state. 
In our case, $\Delta(p)$ becomes $c_x p_x+i c_y p_y$, $c_y\gg c_x$ in the 
long wave length limit $p\rightarrow 0$.

More experimental and theoretical studies are needed to decide the 
symmetry of the gap potential and to understand the underlying physics 
at the half-filled second Landau level. 
To compare the theory with experiments quantitatively, 
effects of the Landau level mixing, finite temperature, 
finite thickness of 2D layer, and inplane magnetic field 
need to be included in the calculation. 
Furthermore, Josephson effect, vortex excitations, 
and edge states in our model are important future subjects. 

\acknowledgements

I would like to thank K. Ishikawa, T. Aoyama and especially J. Goryo for 
useful discussions. 
Part of this work was done when I took part in {\it Tenth International 
Conference on Recent Progress in Many-Body Theories}, Seattle. 
This work was partially supported by the special Grant-in-Aid for 
Promotion of Education and Science in Hokkaido University provided by the 
Ministry of Education, Science, Sport, and Culture, the Grant-in-Aid for 
Scientific Research on Priority area (Physics of CP violation) 
(Grant No. 12014201), and the Grant-in aid for International Science 
Research (Joint Research 10044043) from the Ministry of Education, 
Science, Sports, and Culture, Japan.

\begin{figure}
\epsffile{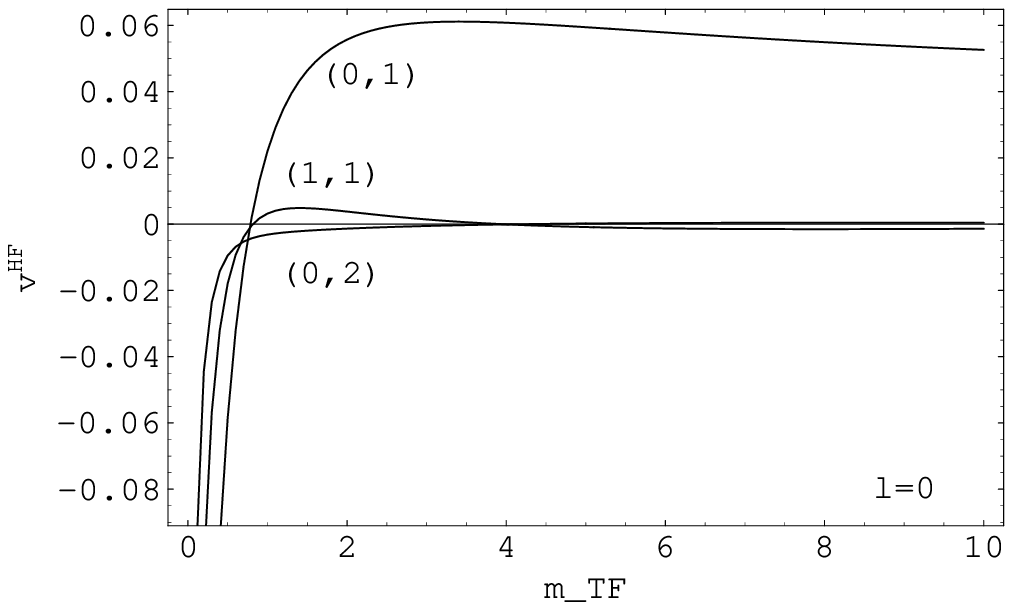}
Fig.~1. The $m_{\rm TF}$ dependence of $v^{\rm HF}(0,1)$, $v^{\rm HF}(1,1)$, 
and $v^{\rm HF}(0,2)$ for $l=0$. 
The unit of energy is $q^2/a$. The same unit is used in all other 
figures. 
\end{figure}
\begin{figure}
\epsffile{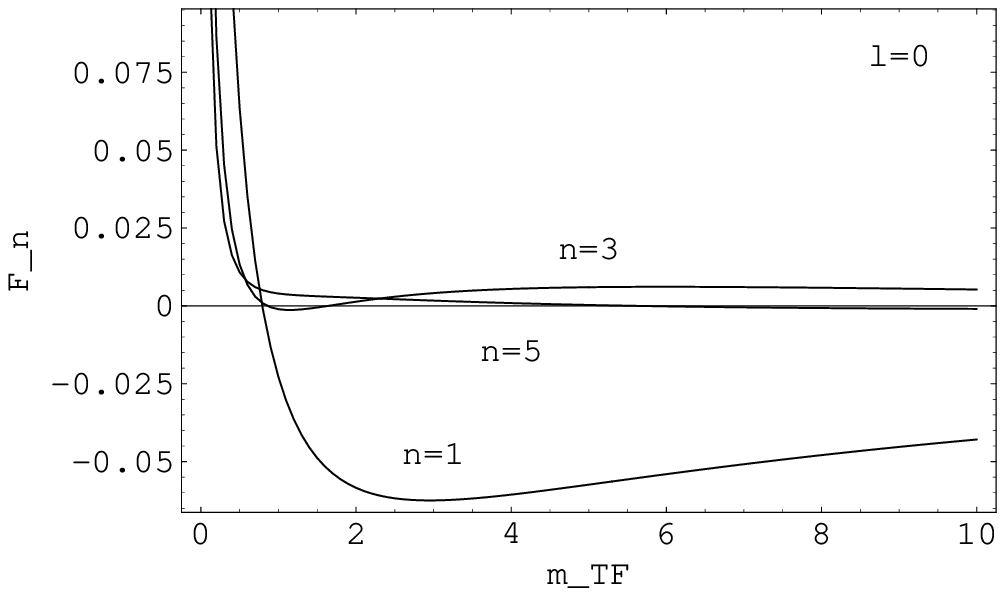}
Fig.~2. The $m_{\rm TF}$ dependence of pseudopotentials 
$F_1$, $F_3$, and $F_5$ for $l=0$. 
\end{figure}
\begin{figure}
\epsffile{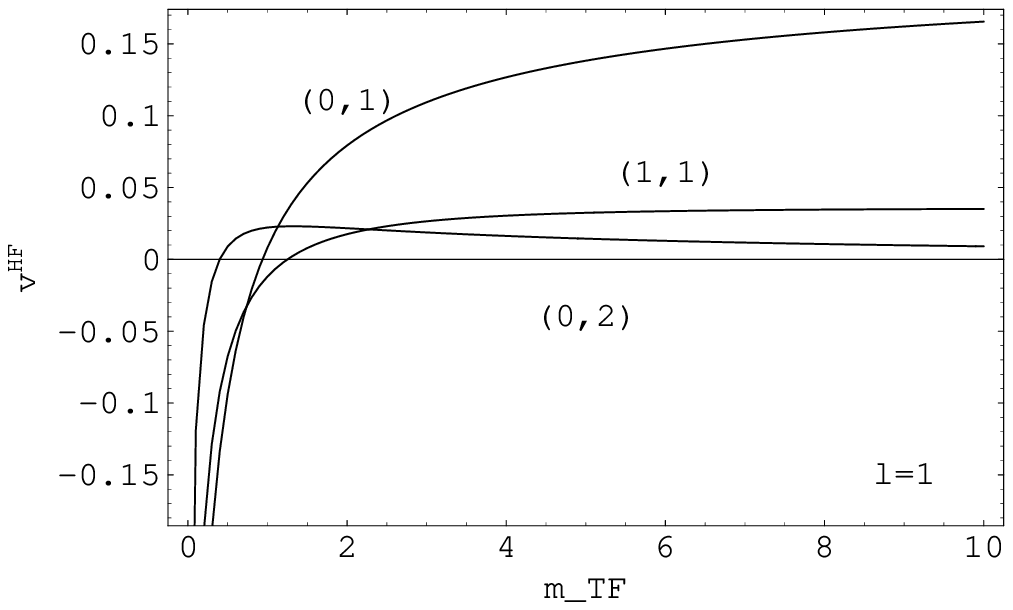}
Fig.~3. The $m_{\rm TF}$ dependence of 
$v^{\rm HF}(0,1)$, $v^{\rm HF}(1,1)$, and $v^{\rm HF}(0,2)$ for $l=1$. 
\end{figure}
\begin{figure}
\epsffile{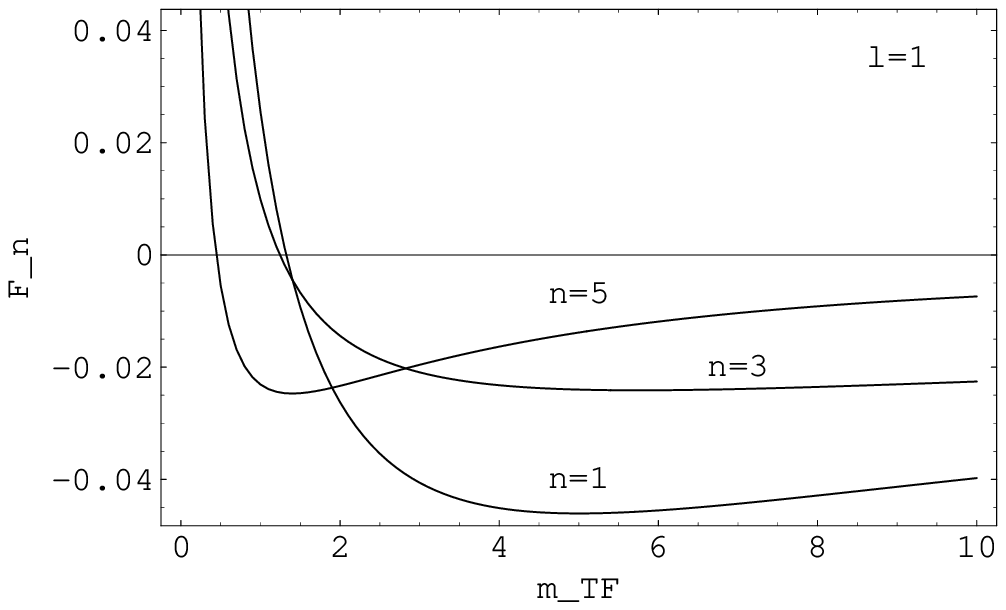}
Fig.~4. The $m_{\rm TF}$ dependence of pseudopotentials 
$F_1$, $F_3$, and $F_5$ for $l=1$. 
\end{figure}
\begin{figure}
\epsffile{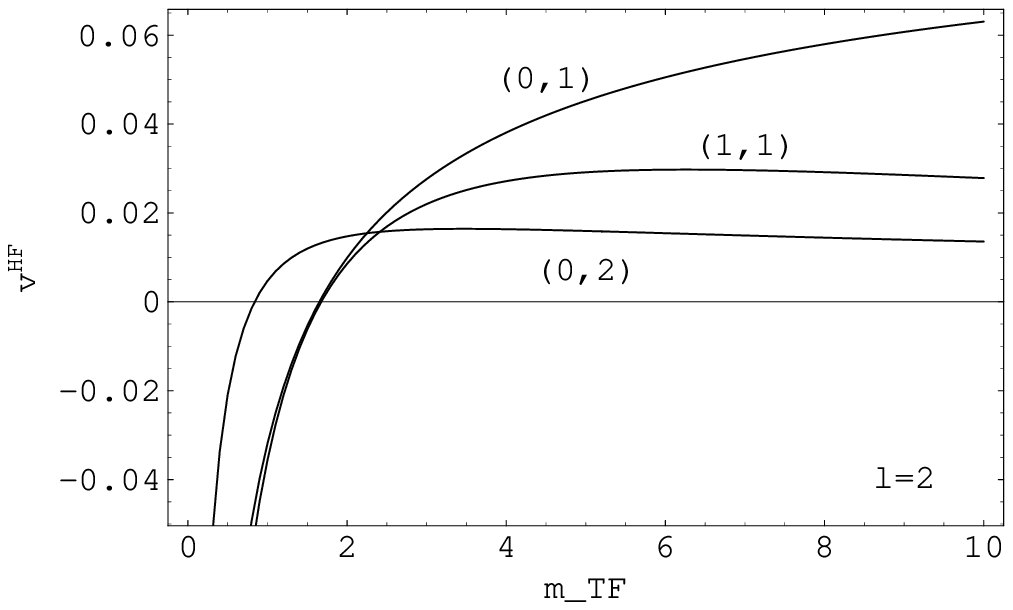}
Fig.~5. The $m_{\rm TF}$ dependence of 
$v^{\rm HF}(0,1)$, $v^{\rm HF}(1,1)$, and $v^{\rm HF}(0,2)$ for $l=2$. 
\end{figure}
\begin{figure}
\epsffile{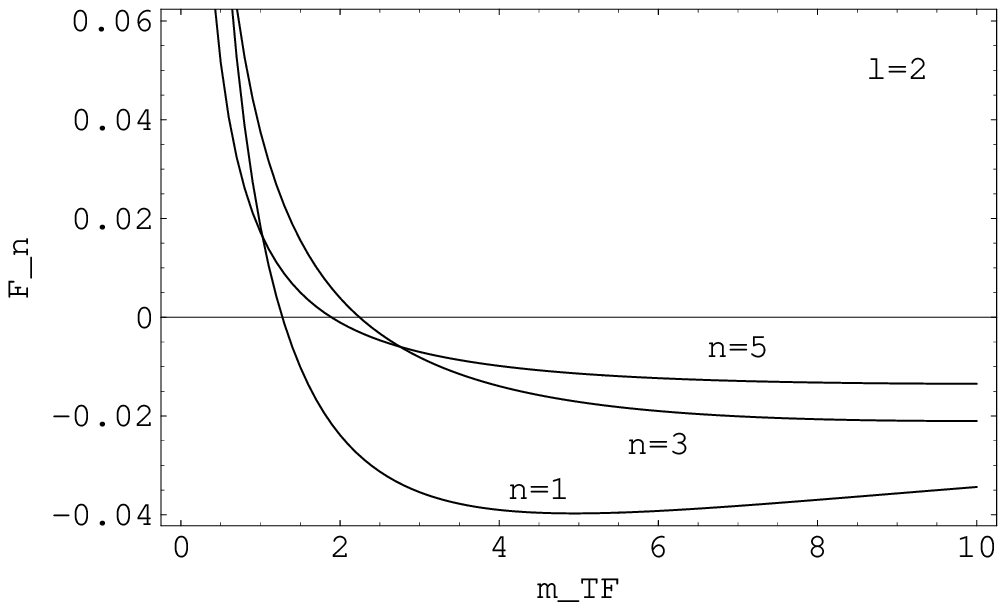}
Fig.~6. The $m_{\rm TF}$ dependence of pseudopotentials 
$F_1$, $F_3$, and $F_5$ for $l=2$. 
\end{figure}
\begin{figure}
\epsffile{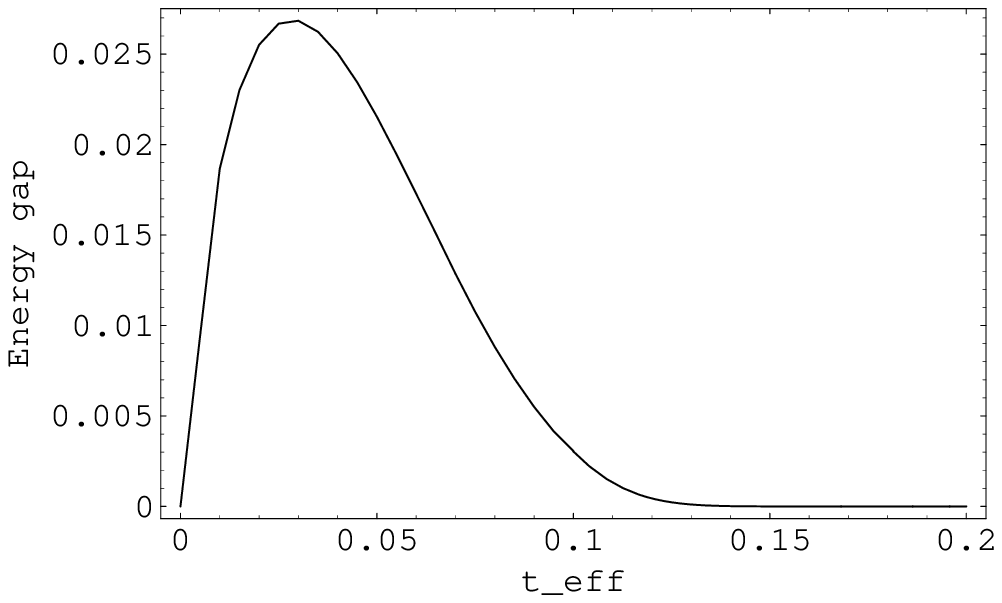}
Fig.~7. The $t_{\rm eff}$ dependence of the energy gap $\Delta E$ 
for $l=1$. 
Crossover occurs at $t_{\rm eff}\approx0.01$. 
The energy gap becomes maximum at $t_{\rm eff}\approx0.03$ and 
vanishes at $t_{\rm eff}\approx0.2$. 
\end{figure}
\begin{figure}
\epsffile{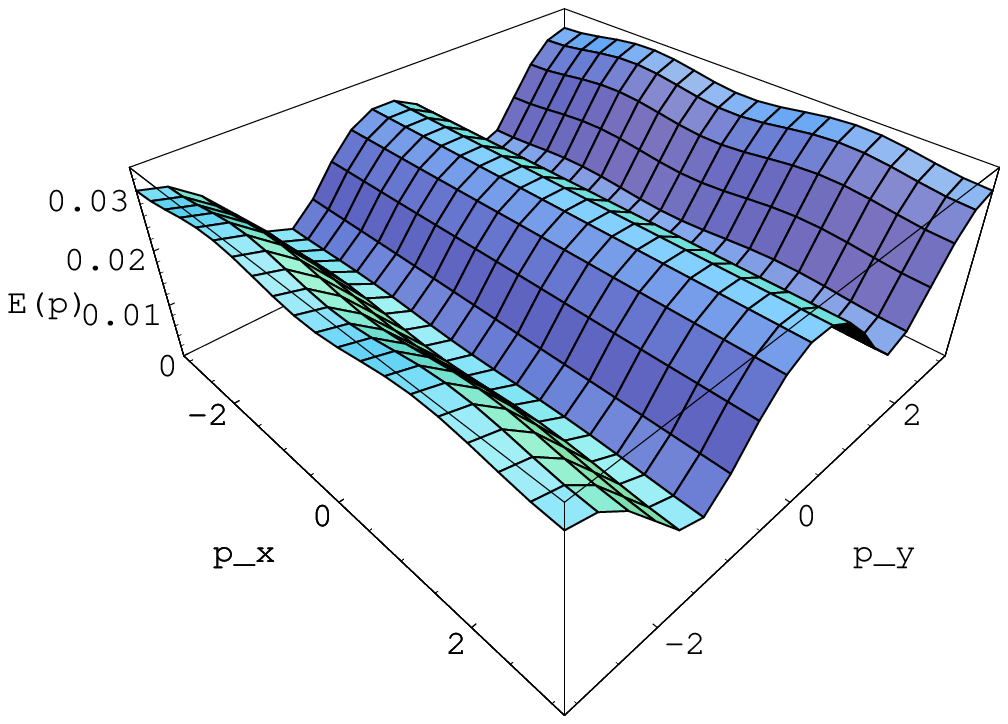}
Fig.~8. Spectrum of the quasiparticle $E(p)$ 
 at $t_{\rm eff}=0.03$ for $l=1$. 
\end{figure}
\begin{figure}
\epsffile{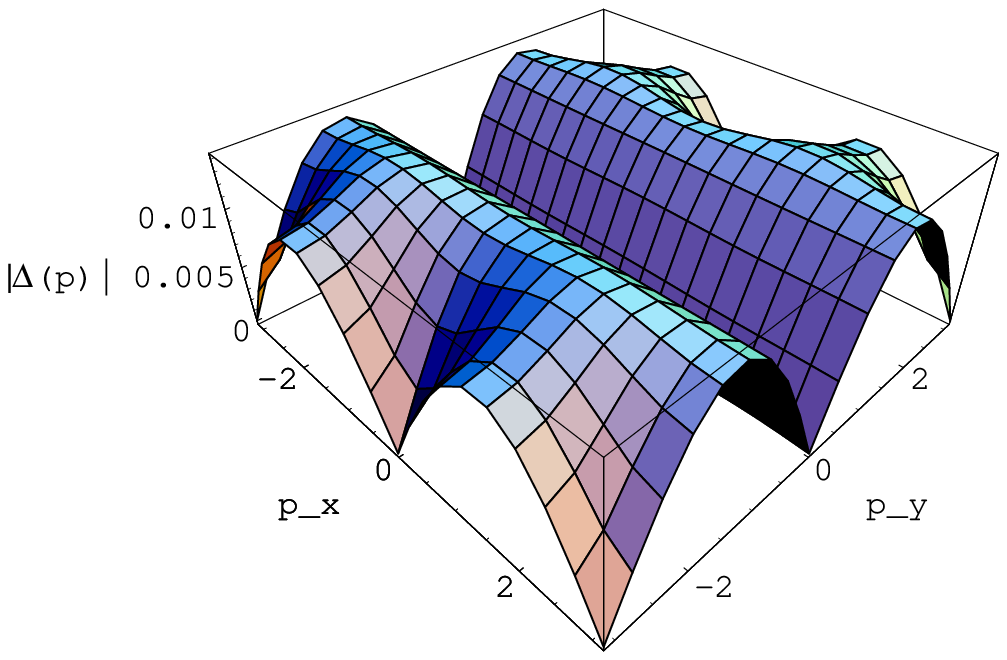}
Fig.~9. Absolute value of the gap potential $\vert\Delta(p)\vert$ at 
$t_{\rm eff}=0.03$ for $l=1$. 
\end{figure}
\begin{figure}
\epsffile{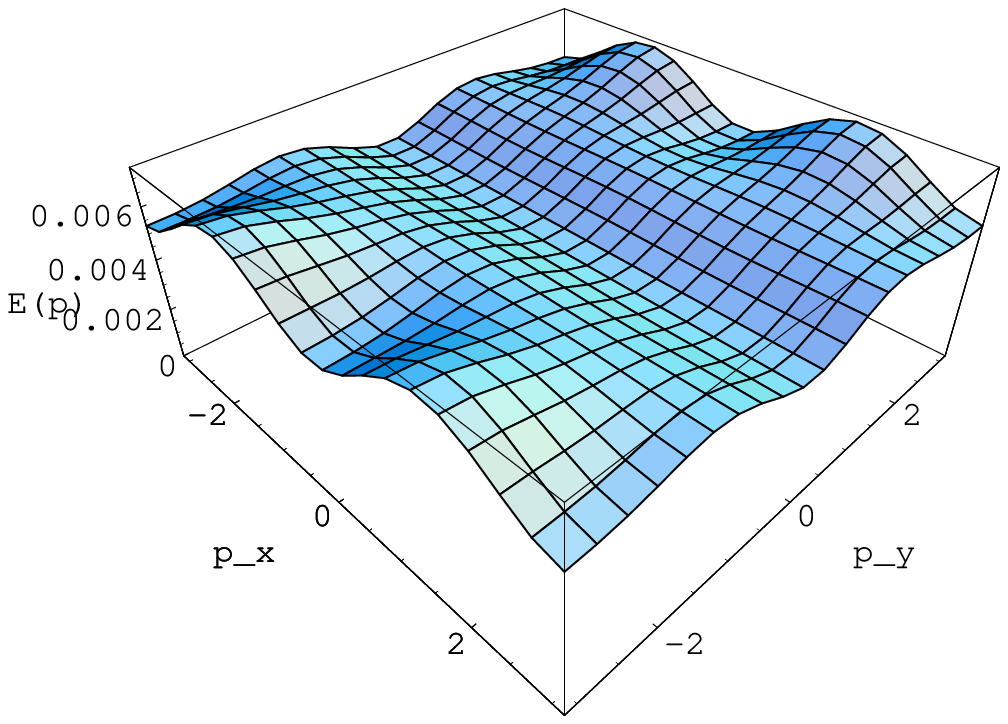}
Fig.~10. Spectrum of the quasiparticle $E(p)$ 
at $t_{\rm eff}=0.005$ for $l=1$. 
\end{figure}

\begin{references}
\bibitem{a} 
D. C. Tsui, H. L. Stormer, and A. C. Gossard, Phys. Rev. Lett. 48 
(1982) 1559.  
\bibitem{b} 
R. B. Laughlin, Phys. Rev. Lett. 50 (1983) 1395. 
\bibitem{c} 
J. K. Jain, Phys. Rev. Lett. 63 (1989) 199; 
B. I. Halperin, P. A. Lee, and N. Read, Phys. Rev. B 47 (1993) 7312; 
{\it Composite Fermion: A Unified View of the Quantum Hall Regime}, edited by 
O. Heinonen (World Scientific, Singapore, 1998). 
\bibitem{d} 
R. L. Willett, J. P. Eisenstein, H. L. Stormer, D. C. Tsui, 
A. C. Gossard, and J. H. English, Phys. Rev. Lett. 59 (1987) 1776; 
W. Pan, J.-S. Xia, V. Shvarts, E. D. Adams, H. L. Stormer, D. C. Tsui, 
L. N. Pfeiffer, K. W. Baldwin, and K. W. West, ibid. 83 (1999) 3530. 
\bibitem{e} 
R. Morf, Phys. Rev. Lett. 80 (1998) 1505. 
\bibitem{f} 
E. H. Rezayi and F. D. M. Haldane, Phys. Rev. Lett. 84 (2000) 4685. 
\bibitem{g} 
G. Moore and N. Read, Nucl. Phys. B 360 (1991) 362; 
 N. Read and G. Moore, Prog. Theor. Phys. Suppl. 107 (1992) 157. 
\bibitem{h} 
M. Greiter and F. Wilczek, Nucl. Phys. B 370 (1992) 577; 
M. Greiter, X.-G. Wen, and F. Wilczek, ibid. 374 (1992) 567; 
Phys. Rev. Lett. 66 (1991) 3205; 
M. Greiter, Prog. Theor. Phys. Suppl. 107 (1992) 145. 
\bibitem{fra}
E. Fradkin, C. Nayak, A. Tsvelik, and F. Wilczek, Nucl. Phys. 
B 516 (1998) 704; E. Fradkin, C. Nayak, and K. Schoutens, ibid., 
546 (1999) 711. 
\bibitem{sing}
F. D. M. Haldane and E. H. Rezayi, Phys. Rev. Lett. 60 (1988) 956, 1886(E). 
\bibitem{i} 
K. Ishikawa, N. Maeda, and T. Ochiai, Phys. Rev. Lett. 82 (1999) 4292. 
\bibitem{j} 
K. Ishikawa, N. Maeda, T. Ochiai, and H. Suzuki, Physica E 4 
(1999) 37; K. Ishikawa and N. Maeda, Prog. Theor. Phys. 97 (1997) 507. 
N. Imai, K. Ishikawa, T. Matsuyama, and I. Tanaka, Phys. Rev. B 42 (1990) 
10610. 
\bibitem{k} 
N. Maeda, Phys. Rev. B 61 (2000) 4766. 
\bibitem{stripe}
M. P. Lilly, K. B. Cooper, J. P. Eisenstein, L. N. Pfeiffer, and 
K. W. West, Phys. Rev. Lett. 82 (1999) 394; 
R. R. Du, D. C. Tsui, H. L. Stormer, L. N. Pfeiffer, K. W. Baldwin, and 
K. W. West, Solid State Commun. 109 (1999) 389. 
\bibitem{o} 
W. Pan, R. R. Du, H. L. Stormer, D. C. Tsui, L. N. Pfeiffer, 
K. W. Baldwin, and K. W. West, Phys. Rev. Lett. 83 (1999) 820; 
M. P. Lilly, K. B. Cooper, J. P. Eisenstein, L. N. Pfeiffer, and 
K. W. West, ibid. 83 (1999) 824.
\bibitem{pere}
A. M. Perelomov, Teor. Mat. Fiz. 6 (1971) 213 
[Theor. Math. Phys. 6 (1971) 156]; 
V. Bargmann, P. Butera, L. Girardello, and J. R. Klauder, 
Rep. Math. Phys. 2 (1971) 221. 
\bibitem{loc}
Strictly speaking, the Wannier basis is weekly localized state. 
That is, it is normalizable, but its standard deviation of position 
is infinite. 
See Ref.~\cite{j}. 
\bibitem{m} 
A. A. Koulakov, M. M. Fogler, and B. I. Shklovskii, Phys. Rev. Lett. 
76 (1996) 499; M. M. Fogler, A. A. Koulakov, and B. I. Shklovskii, 
Phys. Rev. B 54 (1996) 1853; 
R. Moessner and J. T. Chalker, ibid. 54 (1996) 5006. 
\bibitem{n} 
T. Jungwirth, A. H. MacDonald, L. Smarcka, and S. M. Girvin, Phys. Rev. B 
60, 15574 (1999); E. H. Rezayi, F. D. M. Haldane, and K. Yang, 
Phys. Rev. Lett. 83 (1999) 1292; T. Stanescu, I. Martin, and 
P. Phillips, Phys. Rev. Lett. 84 (2000) 1288. 
\bibitem{hal}
F. D. M. Haldane, Phys. Rev. Lett. 51 (1983) 605. 
\bibitem{q} 
J. Goryo and K. Ishikawa, Phys. Lett. A 260 (1999) 294.
\bibitem{will}
R. L. Willett, H. L. Stormer, D. C. Tsui, A. C. Gossard, and 
J. H. English, Phys. Rev. B 37 (1988) 8476. 
\bibitem{l} 
T. Morinari, Phys. Rev. Lett. 81 (1998) 3741.
\bibitem{bon} 
N. E. Bonesteel, Phys. Rev. Lett. 82 (1999) 984.
\bibitem{bas}
G. Baskaran, Physica B 212 (1995) 320. 
\bibitem{dip}
N. Read, Semicond. Sci. Technol. 9 (1994) 1859; 
V. Pasquier and F. D. M. Haldane, Nucl. Phys. B 516 (1998) 719. 
\bibitem{p} 
N. Read and D. Green, Phys. Rev. B 61 (2000) 10267. 
\end{references}
\end{document}